\documentclass[twocolumn,prb]{revtex4-1}
\usepackage{graphicx}    
\usepackage{dcolumn}     
\usepackage{bm}          
\usepackage{amsmath,amssymb}

\usepackage{color}

\begin{document}
\title{Charge/quadrupole fluctuations and gap anisotropy in BiS$_2$-based superconductors}
\author{
Katsuhiro Suzuki$^1$, Hidetomo Usui$^2$, Kazuhiko Kuroki$^2$, Hiroaki Ikeda$^3$}
\affiliation{$^1$ Research Organization of Science and Technology, 
Ritsumeikan University, Kusatsu, Shiga 525-8577, Japan}
\affiliation{$^2$ Department of Physics, 
Osaka University, 
Toyonaka, Osaka 560-0043, Japan}
\affiliation{$^3$ Department of Physics, 
Ritsumeikan University, Kusatsu, Shiga 525-8577, Japan}
\date{\today}
\begin{abstract}
Recent angle-resolved spectroscopy in BiS$_2$-based superconductors has indicated that the superconducting gap amplitude possesses remarkable anisotropy and/or a sign change on a small Fermi pocket around $X$ point. It implies a possibility of an unconventional pairing state. Here we study the gap anisotropy in superconductivity mediated by inherent charge/quadrupole fluctuations in an extended Hubbard model, which includes inter-site interaction between Bi and S atoms. The first-principles downfolded band structure is composed of Bi $6p_x/p_y$ and S $3p_x/p_y$ orbitals on a  BiS$_2$ single layer. Evaluating the linearized gap equation, we find that the ferroic charge/quadrupole fluctuation driven by the inter-site interaction leads to a fully-gapped $d_{x^2-y^2}$-wave pairing state, in which the gap amplitude has sizable anisotropy on the Fermi surface.
\end{abstract}
\pacs{PACS numbers: }
\maketitle
\section{Introduction}
Recently discovered BiS$_2$-based layered superconductors, Bi$_4$O$_4$S$_3$ [\onlinecite{PhysRevB.86.220510}] and \textit{Ln}O$_{1-x}$F$_x$BiS$_2$ (\textit{Ln} = Lanthanide)\cite{JPSJ.81.114725,1.4790322,Ceolin:a13371,doi:10.1021/ja508564s,raeyJha}, have attracted a great interest as the related materials of iron-based superconductors\cite{ja800073m}. The highest transition temperature $T_c=10.6$K is observed in LaO$_{0.5}$F$_{0.5}$BiS$_2$ [\onlinecite{JPSJ.81.114725}]. The parent material LaOBiS$_2$ is semiconducting, and possesses a crystal structure with alternating stacking of BiS$_2$ twin layers and $Ln$O insulating blocking layers. Superconductivity emerges via electron doping by substituting O with F. Owing to the layered structure, the electronic structure is two-dimensional, and the BiS$_2$ twin layers become conductive with electron doping. The electronic band constructing the Fermi surface is mainly composed of the Bi $6p_x$ and $6p_y$ orbitals. Therefore, it is expected that these orbitals have a relatively large spin-orbit coupling\cite{doi:10.7566/JPSJ.84.012001}. Moreover, due to the non-symmorphic space group, the BiS$_2$ twin layers locally break the inversion symmetry at a Bi site. These features, shared with superconductors with a zigzag chain, CrAs [\onlinecite{doi:10.7566/JPSJ.83.093702}] and UCoGe [\onlinecite{TROC1988389}], are also fascinating in terms of non-centrosymmetric superconductors\cite{sigrist}.

Concerning the pairing state and mechanisms, two possibilities, the conventional $s$-wave mediated by the electron-phonon interaction\cite{0295-5075-101-4-47002,PhysRevB.87.115124} and unconventional superconductivity driven by the purely electronic interactions\cite{0295-5075-108-2-27006,PhysRevB.86.220501,PhysRevB.87.081102,Agatsuma201673,PhysRevB.91.024512, arXiv1701.02909}, have been theoretically investigated in the early stage of the study\cite{NSMusui}. Experimentally, there is no strong evidence of the electron correlation effect. 
Measurements of penetration depth and thermal conductivity indicate that NdO$_{0.7}$F$_{0.3}$BiS$_2$ is a fully gapped superconductor\cite{0953-8984-27-22-225701,doi:10.7566/JPSJ.85.073707}. These observations imply that the superconducting pairing mechanism in this system is the conventional phononic mechanism. However, a recent measurement of field-angle dependent Andreev reflection spectroscopy\cite{arXiv1604.06325} has reported that the superconducting gap amplitude is highly anisotropic. Also, angle-resolved photo-emission spectroscopy (ARPES)\cite{Ota} has indicated the presence of remarkable anisotropy and/or a possibility of sign-change of the superconducting gap on a small Fermi pocket around $X$ point. These observations imply a possibility of an unconventional pairing mechanism in this superconductor. In general, such anisotropic gap structure needs an unconventional mechanism, for instance, strongly $k$-dependent fluctuations, or two kind of competitive forces, such as electron-phonon attractive force and electron repulsive force. In addition, the observation of ``checkerboard stripe'' pattern in STM/STS measurements\cite{JPSJ.83.113701} is indicative of the importance of charge/orbital fluctuation.

Here, to clarify this point, we study in detail a gap anisotropy of unconventional superconductivity induced by purely electronic repulsive forces. First of all, we perform the first-principles calculations\cite{Blaha1990399} of LaOBiS$_2$ without the spin-orbit coupling. Next, we construct a downfolded eight-band tight-binding Hamiltonian by using the maximally localized Wannier functions (MLWFs)\cite{MLWF,Kunes20101888}. The target band consists of $6p_x/p_y$ orbitals of two Bi atoms and $3p_x/p_y$ orbitals of two in-plane S atoms in the unit cell. Furthermore, by neglecting small inter-layer hopping integrals, the eight-orbital model is reduced to be the four-orbital model in a single BiS$_2$ layer. We elucidate charge/orbital fluctuations in this four-orbital model for electron doping corresponding to $x=0.3$ within the random phase approximation (RPA). As the purely electronic interactions, in addition to the conventional Hubbard-type on-site Coulomb interactions, we consider inter-site interactions between Bi and S atoms. We find that the inter-site interactions enhance a ferroic charge fluctuation, especially, orbital-dependent inter-site interactions lead to a ferroic quadrupole fluctuation. This may be consistent with the ``checkerboard stripe'' observed in STM/STS\cite{JPSJ.83.113701}. Furthermore, solving the superconducting gap equation, we find the possibility of fully-gapped $d_{x^2-y^2}$-wave ($B_{1g}$) pairing state mediated by such charge/quadrupole fluctuations. The gap amplitude on the Fermi surface has sizable anisotropy, which is similar to the experimental observations. Finally, we realize that the inter-site interactions between Bi and S atoms are the key ingredients to understand the superconductivity of this material, although it may be difficult to understand it in terms of purely electronic interactions.

\section{Model Hamiltonian and Random Phase Approximation}
\begin{figure}[htbp]
\centering
\includegraphics[width=8cm,clip]{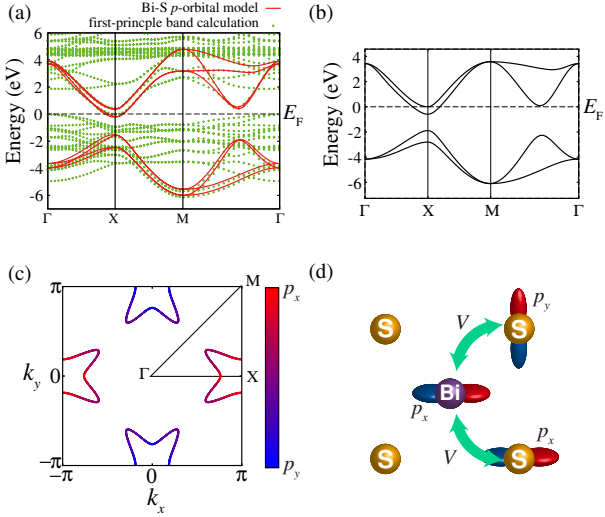}
\caption{(color online) (a) Band structure obtained by the first-principles calculation of LaOBiS$_2$ (green dots) and a downfolded eight-orbital model (red line). (b) Band structure in our four-orbital model, and (c) the Fermi surface colored by the weight of Bi $6p_x$ (red) and $6p_y$ (blue). Here, the electron filling corresponds to $x=0.3$. (d) Schematic diagram of inter-site interaction $V$ between Bi and S.}
\end{figure}

The BiS$_2$-based superconductors have a common feature of two-dimensional Fermi surface, which mainly comes from the Bi $6p$ orbitals. In order to study the characteristic low-energy effective model, we use a downfolded band structure of LaO$_{0.5}$F$_{0.5}$BiS$_2$ as in the previous study\cite{PhysRevB.86.220501}. We start with the first-principles calculations of LaO$_{0.5}$F$_{0.5}$BiS$_2$ using the {\tt WIEN2k} package\cite{Blaha1990399} with the experimental lattice parameters\cite{JPSJ.81.114725}. We take $RK_{max}=7$ and $512$ $k$-points grid, and adopt the GGA-PBE exchange correlation functional\cite{PhysRevLett.100.136406}. Then, we describe the target bands near the Fermi level based on the MLWFs\cite{MLWF,Kunes20101888} of Bi $6p_x/p_y$ and S $3p_x/p_y$ orbitals. Finally, we obtain an effective eight-orbital tight-binding model considering the BiS$_2$ twin layer in the unit cell. It well reproduces the original band structure as shown in Fig.~1~(a). In the obtained transfer integrals, we find that the inter-layer hopping integrals are very small due to the two-dimensional structure. Indeed, we can see in Fig.~1~(b) that the four-orbital model without the inter-layer hoppings\cite{ham}, i.e., the BiS$_2$ single layer model relatively well reproduces the band structure near the Fermi level. Note that the Fermi level has been shifted to the level corresponding to F-doping $x=0.3$, not $x=0.5$. Fig.~1~(c) depicts the corresponding Fermi surface colored by the weight of Bi $6p_x/p_y$ orbitals, where the $x/y$ direction corresponds to a Bi-Bi direction, rotating by 45 degree from $X/Y$ in the previous study\cite{PhysRevB.86.220501}.

Here we consider as usual the Hubbard-type interactions on each atomic site, 
\begin{align}
\begin{split}
\mathcal{H}^\textrm{intra}_{I}=
&\sum_{i}\left[\sum_{\nu}Un_{i\nu\uparrow}n_{i\nu\downarrow}
+\sum_{\mu>\nu}U'n_{i\nu}n_{i\mu}\right.  \\
&\left.+\sum_{\mu>\nu}J\hat{S}_{i\nu}\cdot\hat{S}_{i\mu}
+\sum_{\nu\neq\mu}J'c^\dagger_{i\nu\uparrow}c^\dagger_{i\nu\downarrow}
c_{i\mu\downarrow}c_{i\mu\uparrow}\right],
\end{split}
\end{align}
with
\begin{align*}
& n_{i\nu}=\sum_{\sigma}n_{i\nu\sigma}=\sum_{\sigma}c^{\dagger}_{i\nu\sigma}c_{i\nu\sigma},  \\
& \hat{S}_{i\nu}=\sum_{\alpha\beta}c^{\dagger}_{i\nu\alpha}\hat{\sigma}_{\alpha\beta} c_{i\nu\beta},
\end{align*}
where $\hat{\sigma}$ is the Pauli matrices, and $c_{i\nu\sigma}$ is an annihilation operator of a spin-$\sigma$ electron on $\nu$ orbital ($p_x$ or $p_y$) at $i$ site. For simplicity, we fix the ratio of each interaction to the intra-orbital repulsion $U$ as follows,  $U'=3U/4$ for the inter-orbital interaction, and $J=J'=U/8$ for the Hund's coupling $J$ and the pair hopping $J'$. 
In addition, considering a wide spread of MLWFs of Bi $6p$ orbitals, we include the inter-site interactions $V_\pm=V\pm V'$ between Bi and S atoms as shown in Fig.~1~(d), 
\begin{align}
\mathcal{H}^\textrm{inter}_{I}=\sum_{\langle i,j \rangle}\sum_{\nu \ne \mu}
V_+ n_{i\nu}n_{j\nu}+V_- n_{i\nu}n_{j\mu},
\end{align}
where $\langle i,j \rangle$ denotes a summation for the neighboring Bi and S atoms. Hereafter, $V'=0$ unless otherwise noted.

Now, let us investigate what kinds of fluctuations grow in the extended Hubbard model within the RPA. In the present four-orbital model, the spin and charge (orbital) susceptibilities are evaluated through the following $8\times 8$ matrices,
\begin{align}
\hat\chi_{s(c)}(q)=\hat\chi_0(q)\left(\hat{1}-\hat{\Gamma}_{s(c)}\hat\chi_0(q)\right)^{-1},
\end{align}
where $q=(\bm{q},i\nu_n)$ with boson Matsubara frequencies $\nu_n$, and $\hat{1}$ is an identity matrix. 
Each element of the irreducible susceptibility matrix $\hat\chi_0(q)$ is obtained from
\begin{align}
\chi_0^{12,34}(q)=&-\frac{T}{N}\sum_{k}G^{13}_0(k+q)G^{42}_0(k),
\end{align}
where labels $1-4$ symbolically denote an atom (Bi/S) and its orbital ($p_x/p_y$) in the unit cell, and $G^{13}_0(k)$ is the one-particle bare Green's function between label $1$ and label $3$. 
Moreover, the elements of the bare interaction matrix $\hat\Gamma_{s/c}$ are given by
\begin{align}
\Gamma_s^{12,34}=\left\{
\begin{array}{ll}
S_{\ell_1\ell_2,\ell_3,\ell_4} & (1-4 \in \mbox{Bi})\\
0 & (\mbox{otherwise})
\end{array}\right.
,
\end{align}
\begin{align}
-\Gamma_c^{12,34}=\left\{
\begin{array}{ll}
C_{\ell_1\ell_2,\ell_3,\ell_4} & (1-4 \in \mbox{Bi})\\
V_{\ell_1\ell_2,\ell_3,\ell_4}(\bm{q}) & (1,2 \in \mbox{Bi and } 3,4 \in \mbox{S})\\
V^*_{\ell_1\ell_2,\ell_3,\ell_4}(\bm{q}) & (1,2 \in \mbox{S and } 3,4 \in \mbox{Bi})\\
0 & (\mbox{otherwise})
\end{array}\right.
,
\end{align}
where $\ell_1-\ell_4$ denotes an orbital $p_x/p_y$ on a Bi/S atom. 
The onsite Coulomb repulsions $\hat{S}/\hat{C}$ are as usual given by
$S_{\ell\ell,\ell\ell}=U$, $S_{\ell m,\ell m}=U'$, $S_{\ell\ell,mm}=J$, $S_{\ell m,m\ell}=J'$, and
$C_{\ell\ell,\ell\ell}=U$, $C_{\ell m,\ell m}=2J-U'$, $C_{\ell\ell,mm}=2U'-J$, $C_{\ell m,m\ell}=J'$
with $\ell (m)=p_x/p_y$ and $\ell\neq m$.
The additional inter-site interactions $\hat{V}(\bm{q})$ are expressed by $V_{\ell\ell,\ell\ell}(\bm{q})=2\gamma(\bm{q})V_+$ and $V_{\ell\ell,mm}(\bm{q})=2\gamma(\bm{q})V_-$ from Eq.~(2), where $\gamma(\bm{q})=\sum_{j}\exp(i\bm{q}\cdot\bm{R}_j)$ and $\bm{R}_j$ is a relative coordinate between the neighboring Bi and S atoms. 
In the form of Eq.~(3), the Stoner factors in the spin (charge) sectors, $\alpha_{s(c)}$, are defined as the maximum eigenvalue of $\hat{\Gamma}_{s(c)}\hat\chi_0(q)$. They are measures of the dominant spin (charge) fluctuations. When they equal to unity, the corresponding spin or charge (orbital) ordering can be realized.

Finally, we investigate a possible spin-singlet superconductivity mediated by these dominant fluctuations. For this purpose, we evaluate the linearized gap equation,
\begin{align}
\begin{split}
&\lambda\phi^{56}(\bm{k})=
-\frac{T}{N}\sum_n \sum_{\bm{q}}\sum_{1234}V_s^{51,26}(\bm{q},0) \\
&~~~\times G_0^{13}(\bm{k-q},i\omega_n)\phi^{34}(\bm{k-q})
G_0^{24}(\bm{q-k},-i\omega_n),
\end{split}
\end{align}
with the pairing interaction, 
\begin{align}
\begin{split}
\hat{V}_s(q)&=\frac{1}{2}\left(\hat{C}(q)+\hat{S}\right) \\
&+\frac{3}{2}\hat{S}\hat\chi_s(q)\hat{S}-\frac{1}{2}\hat{C}(q)\hat\chi_c(q)\hat{C}(q).
\end{split}
\end{align}
Here, $\phi^{12}(\bm{k})$ is a superconducting gap function between orbital 1 and 2, and $\lambda$ is the corresponding eigenvalue, which is unity at $T=T_c$. 
With the unitary matrix diagonalizing the four-orbital tight-binding term, $\phi^{12}(\bm{k})$ is transformed into $\Delta(\bm{k})$ in the band representation. 
In the present numerical calculations, we fix $T=0.001$eV, and used $256 \times 256$ $k$-mesh grid and $1024$ Matsubara frequencies.

\section{charge/quadrupole fluctuations and superconductivity}
\subsection{Gap function}
\begin{figure}[htbp]
\centering
\includegraphics[width=8cm,clip]{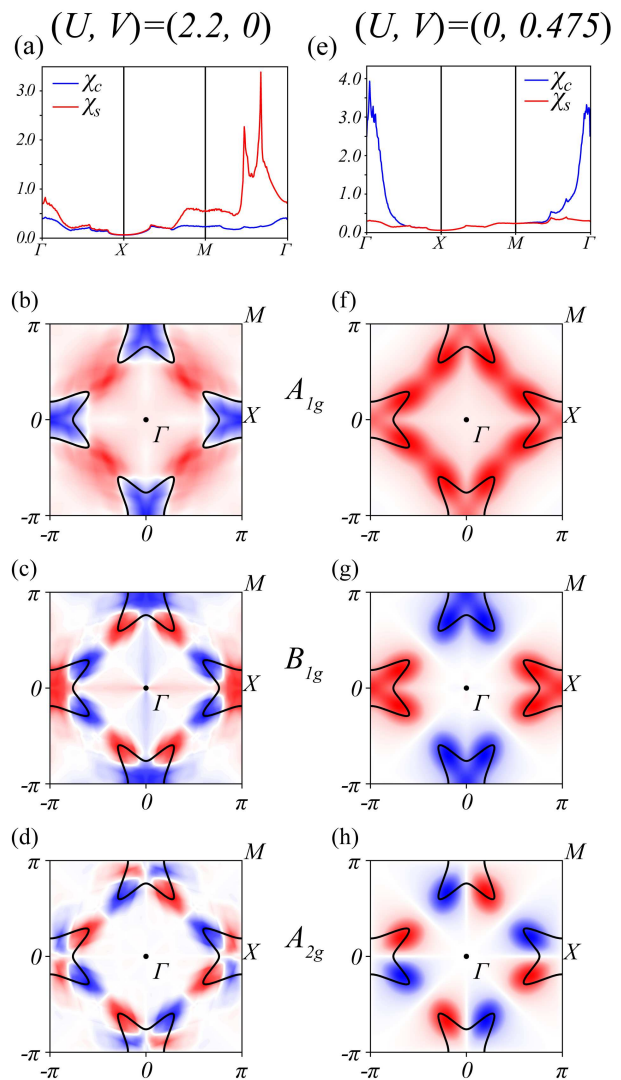}
\caption{(color online) Charge/Spin susceptibility and gap function for each symmetries $A_{1g}$, $B_{1g}$ and $A_{2g}$ at $(U, V)=(2.2, 0.0)$eV (a)-(d) and $(U, V)=(0.0, 0.475)$eV (e)-(h).}
\end{figure}

First, let us discuss the dominant fluctuations and possible gap structure obtained within the RPA. We start with the case of $V=0$ and $U=2.2$eV. 
Fig.~2~(a) depicts the dominant spin/charge fluctuations ($\chi_s^\textrm{max}/\chi_c^\textrm{max}$) along the high-symmetry line. As expected, the magnetic fluctuation $\chi_s$ is enhanced, while the charge fluctuation $\chi_c$ is not enhanced. The characteristic $Q$ structure of $\chi_s^\textrm{max}$ originates from the Fermi surface nesting. From Eqs.~(7) and (8), we calculate possible gap structures in superconductivity mediated by such spin fluctuations. Figs. 2~(b)-(d) indicate $A_{1g}$, $B_{1g}$, $A_{2g}$ gap structures, respectively. The leading pairing state is a $B_{1g}$ state in Fig.~2(c),  and an $A_{2g}$ state in Fig.~2~(d). Eigenvalue $\lambda=1.31$ of the former is larger than $\lambda=1.17$ of the latter. The sequence can be easily changed, depending on the electron filling, as already reported in the previous study\cite{0295-5075-108-2-27006}. Thus, these superconducting states nearly degenerate.

Next, let us consider the case of $(U, V)=(0.0, 0.475)$eV. We illustrate the dominant fluctuations in Fig.~2~(e), and possible gap structures in Figs.~2~(f)-(h). In this case, the spin fluctuations are not enhanced. The dominant fluctuation is a ferroic charge fluctuation. Since the enhanced charge fluctuation favors an isotropic gap on the small Fermi pocket, the fine structure observed in Figs.~2~(b)-(d) are completely or partly lifted, and then, we obtain simple gap structures, the fully gapped $s$-wave $A_{1g}$ state in Fig.~2~(f), the $d_{x^2-y^2}$-wave $B_{1g}$ state in Fig.~2~(g) and the $g_{xy(x^2-y^2)}$-wave $A_{2g}$ state in Fig.~2~(h). Interestingly, due to the smallness of the Fermi pocket, the $d_{x^2-y^2}$-wave $B_{1g}$ state is fully gapped, and the $g_{xy(x^2-y^2)}$-wave $A_{2g}$ state has $d_{xy}$-type line nodes on the Fermi surface. The leading pairing state is $s$-wave $A_{1g}$ state with $\lambda=1.15$. However, with a small but finite $U$, the leading pairing state becomes the fully-gapped $d_{x^2-y^2}$-wave $B_{1g}$ state. The eigenvalue $\lambda=0.94$ for the $B_{1g}$ state is slightly larger than $\lambda=0.82$ for the $A_{2g}$ state due to the presence of small repulsive interaction developing around $Q=(\pi/2,0)$ and the equivalent $Q$ vectors (not shown). 

We realize that all gap functions show bright spots at the corner of the Fermi pocket. Indeed, we can see below in Fig.~4~(a) that the gap amplitude on the Fermi surface is enhanced at around $\theta\sim 45^\circ$, which corresponds to strong suppression of the Fermi velocity in Fig.~4~(c). It is reasonable from the viewpoint of the condensation energy, since the superconducting gap amplitude is large on $k$ points with the high density of states, i.e., small Fermi velocity. 

\subsection{Phase diagram}
In Fig.~3~(a), we show eigenvalues as a function of $V$ for several $U$ along with $\alpha_c$, which is a measure of the Stoner factor for the charge susceptibility $\chi_c$. We can see that when $\alpha_c$ is enhanced as increasing $V$, eigenvalues $\lambda$ are also enhanced and greater than $1$ in a close proximity to the phase boundary of charge density wave (CDW) at $\alpha_c=1$. The leading pairing state is an $s$-wave $A_{1g}$ state at $U=0.0$, but a $d_{x^2-y^2}$-wave $B_{1g}$ state for finite $U$. As increasing $U$, the difference between $A_{1g}$ and $B_{1g}$ shrinks, and then these are nearly degenerate at $U=1.6$. Since these fully-gapped states have almost the same gap structure on the Fermi surface except for the sign, it is reasonable that these are nearly degenerate. For $U>1.6$, the dominant fluctuation changes from the ferroic charge fluctuation into the incommensurate spin fluctuation. The leading pairing state is nodal $d_{x^2-y^2}$ wave state in Fig.~2~(c). These are summarized in the phase diagram of Fig.~3~(b). We conclude that the fully-gapped superconductivity driven by the ferroic charge fluctuation appears near the CDW phase boundary, while the magnetically-driven nodal $d_{x^2-y^2}$-wave state appears near the SDW phase boundary. A broken line indicates $\lambda=1$. 

\begin{figure}[htbp]
\centering
\includegraphics[width=6.5cm,clip]{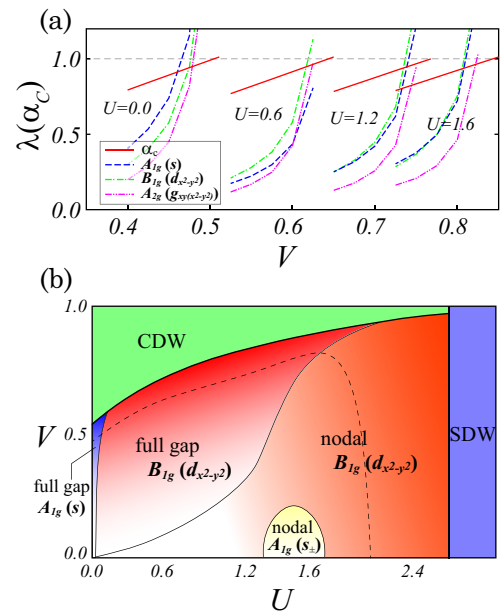}
\caption{(color online) (a) Stoner factor $\alpha_C$ and eigenvalue $\lambda$ of each pairing symmetry as a function of $V$ for several $U$. (b) $U-V$ phase diagram. The broken line corresponds to a phase transition line of $\lambda=1$.}
\end{figure}

\begin{figure}[hbtp]
\centering
\includegraphics[width=8cm,clip]{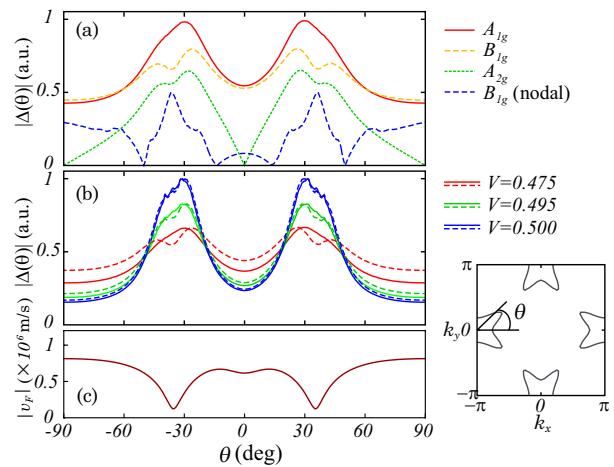}
\caption{(color online) (a) Superconducting gap $|\Delta(\theta)|$ on the Fermi surface. The angle $\theta$ is indicated in the right bottom inset. (b) Development of gap anisotropy in the fully-gapped $A_{1g}$ and $B_{1g}$ states at $U=0$. (c) Anisotropy of the Fermi velocity $|v_F(\theta)|$.}
\label{fig4}
\end{figure}
%
\subsection{Gap anisotropy}Recent experimental observations have implied the strong gap anisotropy in this material\cite{Ota}. Here, let us dissect the gap anisotropy on the Fermi surface for the obtained gap structures. Fig.~\ref{fig4}(a) depicts the angle dependence of gap amplitude $|\Delta(\theta)|$ on the Fermi surface for possible gap structures. The $A_{2g}$ state in Fig.~2(h), which does not appear in the phase diagram of Fig.~3(b), has $d_{xy}$-like symmetry-protected nodes. The fully-gapped $A_{1g}$/$B_{1g}$ state in Fig.~2(f)/(g) also has $d_{xy}$-like anisotropy, although the gap at $\theta=0^\circ$ and $\pm 90^\circ$ is a finite gap minima, not a gap zero.  As indicated in Fig.~\ref{fig4}(b), such anisotropy develops in the close proximity to a CDW phase boundary. The nodal $B_{1g}$ in Fig.~2(c) has fine structure, where the nodal positions are located at $\theta\sim\pm 15^\circ$ and $\pm 50^\circ$. 
Experimentally, the recent ARPES data shows a gap node/minimum at $\theta=0$, but its data is scattered at around $\theta=90$. Then, at least, the behavior at around $\theta=0$ is consistent with the $d_{xy}$-like gap anisotropy. However, the symmetry-protected nodes in $A_{2g}$ state, which cannot be easily lifted, are incompatible with the fully-gapped nature reported by some experiments\cite{0953-8984-27-22-225701,doi:10.7566/JPSJ.85.073707}. Thus, the fully-gapped $A_{1g}$ or $B_{1g}$ state is a possible gap structure in this system. In particular, the latter fully-gapped $d_{x^2-y^2}$-wave $B_{1g}$ state is stable in the wide range of the phase diagram. This state is mediated by the ferroic charge fluctuation, which is driven by the inter-site interactions between Bi and S atom. 

\begin{figure}[htbp]
\centering
\includegraphics[width=8cm,clip]{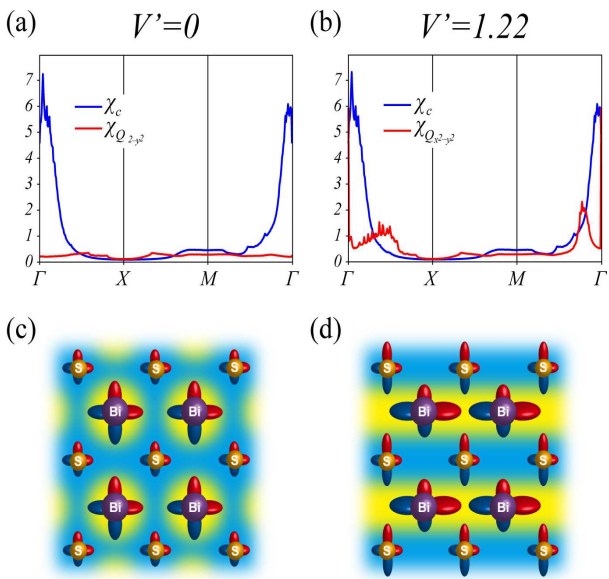}
\caption{(color online) Charge/Quadrupole susceptibilities $(\chi_C/\chi_Q)$ at $(U, V, V')=(0.0, 0.475, 0)$eV (a) and  $(U, V, V')=(0.0, 0.475, 1.22)$eV (b).  Schematic charge distribution of the (c) ferroic charge ordering and (d) quadrupole ordering.}
\end{figure}
%
\subsection{Charge/quadrupole ordering}
Finally, let us discuss possible charge/quadrupole ordering. As mentioned above, the inter-site interaction $V$ between Bi and S atoms lead to the ferroic charge fluctuation. As indicated in Fig.~5~(a), its net component is the charge (electric monopole) fluctuation, defined by $\chi_c=\sum_l\chi_{llll}+\sum_{l\neq m}\chi_{llmm}$. In general, the inter-site interactions are orbital-dependent, that is to say, $V'$ is finite. As indicated in Fig.~5~(b), with increasing $V'$, the $Q_{22}$-type quadrupole fluctuation, $\chi_Q=\sum_l\chi_{llll}-\sum_{l\neq m}\chi_{llmm}$, is enhanced, and then the fully-gapped $B_{1g}$ state is more stable (not shown). The corresponding order is a stripe-type orbital ordering as illustrated in Fig.~5~(d). This orbital ordering may correspond to ``checkerboard stripe'' charge order,  observed by STM/STS\cite{JPSJ.83.113701}. Note that $V_-=V-V'<0$ in this region. It implies that the inter-site attractive force may be important in the emergence of ``checkerboard stripe'' charge order. Therefore, it may be difficult to understand it in terms of purely electronic interactions.

\section{conclusion}
In the present study, we studied the superconducting gap anisotropy in the BiS$_2$-based superconductors. We constructed the first-principles downfolded band structure on the basis of Bi $6p_x/p_y$ and S $3p_x/p_y$ orbitals on a BiS$_2$ single layer. 
In the extended Hubbard model with the inter-site interactions between Bi and S atoms, we found that the ferroic charge/quadrupole fluctuation can be enhanced. This may be related to the observation of ``checkerboard stripe'' charge order. Such charge/quadrupole fluctuation leads to the fully-gapped $d_{x^2-y^2}$-wave pairing state. The obtained gap amplitude has $d_{xy}$-like anisotropy on a Fermi surface, although the gap at $\theta=0^\circ$ and $\pm 90^\circ$ is a finite gap minima, not a gap zero. Such anisotropy is partially consistent with the recent experimental observations. These results indicate that the inter-site interactions are the key ingredients to understand the superconductivity in this system. 

\section{acknowledgment}
We acknowledge Y. Ota, K. Okazaki, S. Shin for the recent ARPES data, and thank T. Nomoto, K. Hattori for valuable comments. This work was partly supported by JSPS KAKENHI Grant No.16H04021, 16H01081, 15H05745, 15H02014, and 25009605. 
\bibliography{BiB}
\end{document}